\documentclass[a4paper,11pt]{article}

\usepackage[dvips]{color,graphicx}

\definecolor{red  }{rgb}{1,0,0}
\definecolor{blue }{rgb}{0,0,1}
\definecolor{green}{rgb}{0,1,0}

\newcommand{\be}{\begin{equation}}
\newcommand{\ee}{\end{equation}}
\newcommand{\bea}{\begin{eqnarray}}
\newcommand{\eea}{\end{eqnarray}}
\newcommand{\ena}{\end{eqnarray}}

\def\half{\frac{1}{2}}

\usepackage{graphicx,amssymb,amsmath,bm,latexsym}
\makeatletter
\@addtoreset{equation}{section}

\makeatother
\newcommand{\vs}[1]{\vspace{#1 mm}}
\newcommand{\hs}[1]{\hspace{#1 mm}}
\renewcommand{\a}{\alpha}
\renewcommand{\b}{\beta}
\renewcommand{\c}{\gamma}
\renewcommand{\d}{\delta}
\newcommand{\e}{\epsilon}

\newcommand{\pa}{\partial}
\newcommand{\nn}{\nonumber\\}
\newcommand{\p}[1]{(\ref{#1})}

\newcommand{\td}{\tilde d}

\begin{document}

\begin{titlepage}

\begin{flushright}
KU-TP 040 \\
\end{flushright}

\vskip .5in

\begin{center}

{\Large\bf Supersymmetric Intersecting Branes on the Waves}
\vskip .5in

{\large
Kei-ichi Maeda,$^{a,b,}$\footnote{E-mail address: maeda``at"waseda.jp}
Nobuyoshi Ohta,$^{c,}$\footnote{E-mail address: ohtan``at"phys.kindai.ac.jp}
Makoto Tanabe$^{b,}$\footnote{E-mail address:
tanabe``at"gravity.phys.waseda.ac.jp}
and
Ryo Wakebe$^{a,}$\footnote{E-mail address:
wakebe``at"gravity.phys.waseda.ac.jp}
}\\
\vs{10}
$^a${\em Department of Physics, Waseda University,
Shinjuku, Tokyo 169-8555, Japan} \\
$^b${\em Waseda Research Institute for Science and Engineering,  Shinjuku, Tokyo 169-8555, Japan}\\
$^c${\em Department of Physics, Kinki University,
Higashi-Osaka, Osaka 577-8502, Japan} \\

\vskip .2in \vspace{.3in}

\begin{abstract}

We construct a general family of supersymmetric solutions in
time- and space-dependent wave backgrounds in general supergravity theories
describing single and intersecting $p$-branes embedded into
time-dependent dilaton-gravity plane waves of an arbitrary (isotropic) profile,
with the brane world-volume aligned parallel to the propagation direction of
the wave. We discuss how many degrees of freedom we have in the solutions.
We also propose that these solutions can be used to describe higher-dimensional
time-dependent ``black holes", and discuss their property briefly.

\end{abstract}

\end{center}

\vfill

\end{titlepage}

\section{Introduction}

The understanding of the fundamental nature and quantum properties
of spacetime is one of the most important questions in theoretical
physics. An example of such problems is the spacetime singularities
that general relativity predicts especially in time-dependent setting.
However, time-dependent solutions are rather difficult subject in the effective
string theories~\cite{time1}-\cite{OPS}.
Dilaton-gravity plane waves provide a rare example of tractable strongly curved
(possibly singular) time-dependent space-time backgrounds.
They also allow a formulation of (time-dependent) matrix theories of quantum
gravity~\cite{bb1}-\cite{MOTW}.
Other time-dependent brane solutions, though non-supersymmetric, are discussed
in~\cite{KU,MOU}.

Brane supergravity solutions also plays important role~\cite{PT}-\cite{MO},
since they lead to the formulation of the AdS/CFT correspondence.
Hence, it is important to derive supergravity $p$-branes embedded
into dilaton-gravity plane waves.

In a recent paper~\cite{CDEG}, the simplest of these solutions which are
supersymmetric configurations corresponding to time-dependent extremal $p$-branes aligned
along the propagation direction of the plane wave were obtained generalizing
earlier work~\cite{OPS,OP}. These solutions are restricted to single brane
solutions. It has been known that this class of solutions can be
extended to much more general intersecting brane configurations,
and it would be interesting to see if the solutions can be generalized to
such general configurations.

In our previous paper~\cite{MOTW}, we have given closely related solutions
for intersecting branes with wave, but we gave supersymmetric solutions without
wave as examples of our solutions. This actually restrict possible solutions
considerably. In this paper we revisit this class of solutions, and
show that our previous solutions actually give a very general family of
solutions which are wider than those for single brane in Ref.~\cite{CDEG}.
Our solutions involve many arbitrary functions and we examine how many
degrees of freedom we have.
We also discuss some physical applications of these solutions to
black hole physics. In particular, we consider intersecting D1-D5 brane system,
and show that we can effectively compactify it to five dimensions
though close to the possible horizon six-dimensional nature of the solution
reappears.

This paper is organized as follows. In the next section, we briefly summarize
our solutions derived in our previous paper~\cite{MOTW}, where we set wave
profile to zero. This gave a strong restriction on the solutions. However,
here we show that relaxing this condition we can obtain much more general
solutions including those in Ref.~\cite{CDEG}.
In sect.~3, we discuss examples of single brane and two intersecting brane
solutions, and use coordinate reparametrization to count the number of
arbitrary functions involved in the solutions.
This shows that our solutions are quite general one.
In sect.~4, we discuss D1-D5 system as an example. We show that this can
be effectively compactified to five dimensions by using periodic functions
in one of the light-like coordinates. Still we show that we can avoid closed
time-like curve by choosing suitable parameters in the compactification.
This produces a black hole system which looks like a fluctuating
black hole. Close to the horizon, the solution exhibits six-dimensional
nature as is usual for any compactified theory.
We calculate Ricci scalar curvature and Kretschmann invariant of this black hole
and find that there is no curvature singularity at the ``horizon'' in six dimensions.
There remain several interesting questions in this kind of solutions,
but we only mention some of these, leaving detailed study for future.
The final section is devoted to concluding remarks.

\section{Time-dependent brane system in supergravity}

The low-energy effective action for the supergravity system
coupled to dilaton and $n_A$- form field strength is given by
\bea
I = \frac{1}{16 \pi G_D} \int d^D x \sqrt{\mathstrut-g} \left[
 R - \frac12 (\pa \Phi)^2 - \sum_{A=1}^m \frac{1}{2 n_A!} e^{a_A \Phi}
 F_{n_A}^2 \right],
\label{action}
\ena
where $G_D$ is the Newton constant in $D$
dimensions and $g$ is the determinant of the metric. The last term
includes both RR and NS-NS field strengths, and $a_A = \frac12
(5-n_A)$ for RR field strength and $a_A = -1$ for NS-NS 3-form.
In the eleven-dimensional supergravity, there is a four-form and no dilaton.
We put fermions and other background fields to be zero.

We take the following metric:
\bea
ds_D^2 = e^{2 \Xi(u,r)} \left[-2dudv + K(u, r) du^2\right] +
\sum_{\a=1}^{d-2} e^{2 Z_\a(u,r)} (dy^\a)^2 +e^{2B(u,r)}\left(dr^2 + r^2
d\Omega_{\td +1}^2\right),
\label{met}
\ena
where $D=d+\tilde d+2$, the coordinates $u$, $v$ and $y^\a, (\a=1,\ldots, d-2)$
parameterize the $d$-dimensional worldvolume  where
the branes belong, and the
remaining $\tilde d + 2 $ coordinates $r$ and angles
are transverse to the brane worldvolume,
$d\Omega_{\tilde d+1}^2$ is the line element of the $(\td+1)$-dimensional
sphere. Note that $u$ and $v$ are null coordinates.
The metric components $\Xi, Z_\a, B$ and the dilaton $\Phi$ and $K$ are assumed
to be functions of $u$ and $r$.
In our previous paper~\cite{MOTW}, we took $K$ depending on $y_\a$ as well,
but this dependence is dropped for simplicity.
For the field strength backgrounds, we take
\bea
F_{n_A} = E_A'(u,r) \, du \wedge dv \wedge dy^{\a_1} \wedge \cdots \wedge
dy^{\a_{q_A -1}} \wedge dr,
\label{eleb}
\ena
where $n_A = q_A +2$.
Throughout this paper, the dot and prime denote derivatives with
respect to $u$ and $r$, respectively.
The ansatz~\p{eleb} means that we have an electric background.
We could, however, also include magnetic background
in the same form as the electric one.

In our previous paper~\cite{MOTW}, we have shown that the solutions to the field
equations are given by
\bea
ds_D^2 &=& \prod_B H_B^{\frac{2 (q_B+1)}{\Delta_B}}
\Bigg[ e^{2\xi(u)}
\prod_A H_A^{-  \frac{2(D-2)}{\Delta_A}} \left(-2dudv +K(u,r) du^2\right)
\nn
&& \hs{20} + \; \sum_{\a=1}^{d-2}
\prod_A H_A^{-  \frac{2\c_A^{(\a)}}{\Delta_A}}
e^{2\zeta_{\a}(u)} (dy^\a)^2 +
e^{2\beta(u)}\left(dr^2 + r^2 d\Omega_{\tilde d+1}^2\right) \Bigg], \nn
&& E_A = \sqrt{\frac{2(D-2)}{ \Delta_A}} H^{-1}_A, \quad
\Phi =\sum_{A}  \frac{\epsilon_A a_A (D-2)}{\Delta_A} \ln H_A + \phi(u),
\label{res1}
\ena
where $H_A$ is a harmonic function
\bea
H_A = h_A(u) + \frac{Q_A}{r^{\td}},
\label{harm1}
\ena
with $h_A$ being an arbitrary function of $u$ and $Q_A$ a constant,
$\e_A= +1 (-1)$ is for electric (magnetic) backgrounds
and
\bea
\c_A^{(\a)} = \left\{ \begin{array}{l}
D-2 \\
0
\end{array}
\right. \hs{5} {\rm for} \hs{3} \left\{
\begin{array}{l}
y^\a \mbox{   belonging  to $q_A$-brane} \\
{\rm otherwise}
\end{array}
\right.
\label{gamma}
\,.
\ena
We have two constraints still to be satisfied:
\bea
\e_A a_A \phi
+ 2 \sum_{\a \in\kern-0.4em / q_A} \zeta_{\a} + 2 \td \beta=0,
\label{cond1}
\ena
\bea
\left(
r^{\tilde{d}+1}K'
\right)'
=-2e^{-2(\xi-\beta)}r^{\tilde d+1}\prod_A H_A^{2(D-2)/\Delta_A}
\left[W(u,r)+V(u)\right]
\label{last_eq}
\,,
\ena
where
\bea
&& W(u,r)\equiv
\sum_{A,B} \frac{(D-2)^2}{\Delta_A \Delta_B} \Big(\frac{\Delta_A}{D-2} \d_{AB}+2\Big)
(\ln H_A)^{\bm\cdot} (\ln H_B)^{\bm\cdot}
+2 \sum_A \frac{D-2}{\Delta_A}(\ln H_A)^{\bm\cdot\bm\cdot}
\nn
&& \hs{20} +\; 4(D-2)(\dot \beta -\dot \xi)
\sum_A\frac{(\ln H_A)^{\bm\cdot}}{\Delta_A}
\,,
\label{def_W}
\\
&&
V(u) \equiv
\sum_{\a=1}^{d-2} \left(\ddot \zeta_{\a} + {\dot \zeta_{\a}}^2\right)
+ (\tilde d + 2)\left(\ddot \beta + {\dot \beta}^2\right)
- 2 \dot \xi \left[
\sum_{\a=1}^{d-2}
 \dot \zeta_{\a} + (\tilde d + 2) \dot \beta
\right]
+ \frac12 (\dot \phi)^2
\,,
\label{def_V}
\ena

In our previous paper~\cite{MOTW}, we first chose $K$ and then solved
Eq.~(\ref{last_eq}). This gave rather strong constraints on possible solutions.
However we can obtain more general solutions if
Eq.~(\ref{last_eq}) is regarded as the equation for $K$ when other metric
functions are given, which is an elliptic type differential equation with
respect to $r$.
Here we generalize our previous solutions~\cite{MOTW} in this viewpoint.
This approach has recently been taken in Ref.~\cite{CDEG} for a single brane.
Our solutions here include single and intersecting brane solutions as well
as wider solutions including more arbitrary functions than those in \cite{CDEG}.

\section{Solutions with time-dependent harmonic functions}

In this section, we present nontrivial solutions with both
$r$- and $u$-dependent harmonic functions $H_A$ in \p{harm1}.

Before presenting our solutions,
we discuss gauge freedom of null-coordinate transformation.
Under the coordinate transformation
\bea
u=X(\tilde u) \,,\quad
v=\tilde v +Y(\tilde u)\,,
\label{coord}
\ena
where $X$ and $Y$ are arbitrary functions of $u$,
we recover the same solution  (\ref{res1}), (\ref{harm1}),
(\ref{cond1}), and (\ref{last_eq}),
by replacing $K$ and $\xi$ with
\bea
\tilde K(\tilde u, r)
&=& K(X(\tilde u), r){d X\over d \tilde u}
-2{d Y(\tilde u)\over d\tilde u}\,.
\label{wave_new}
\\
\tilde \xi(\tilde u)&=&\xi(X(\tilde u))+{1\over 2}\ln
\left({d X\over d\tilde u}\right)
\,,
\ena
respectively.
Using $X$ and $Y$, we can gauge away $\xi$ and a function of $u$ in $K$
 in our solutions.
We will discuss more details of this procedure in the concrete examples shortly.

For all branes in M-theory and superstring theories, we have the
following relation
\bea
\frac{2(D-2)}{\Delta_A}=1
\,.
\ena
In what follows, we assume this relation.
In the following subsections, we present concrete examples of
a single-brane and two intersecting-brane systems.

\subsection{Single brane}

We first consider a single $A$-brane.
In this case, \p{last_eq} gives
\bea
\left(
r^{\tilde{d}+1}K'
\right)'
=-2e^{-2(\xi-\beta)}\left[
r^{\tilde d+1}\left(\ddot{h}_A+2(\dot{\beta}-\dot{\xi})\dot{h}_A+V(u)h_A\right)
+Q_AV(u) r \right]
\,,
\label{l2}
\ena
upon substituting \p{harm1} and sorting out the terms in the orders of $r$.
We can integrate \p{l2} to obtain
\bea
K(u,r)=
{A_1(u)\over 2(\tilde d+2)}r^2-{B_1(u)\over 2(\tilde d-2)}r^{-(\tilde d-2)}
-{C_1(u)\over \tilde d}r^{-\tilde d}+D_1(u)
\,,
\label{l21}
\ena
where
\bea
A_1(u)&=&-2e^{-2(\xi-\beta)}
\left(\ddot{h}_A+2(\dot{\beta}-\dot{\xi})\dot{h}_A+V(u)h_A\right)
\\
B_1(u)&=&-2e^{-2(\xi-\beta)}Q_A V(u)
\ena
and
$C_1(u)$ and  $D_1(u)$ are  arbitrary functions of $u$.

In the present general solutions, we have $(d+4)$ arbitrary functions:
the metric functions $\xi(u), \zeta_\a (u), \beta (u)$, the dilaton field
$\phi(u)$,
 $h_A(u)$, $C_1(u)$ and $D_1(u)$.
We also have one constraint \p{cond1} for those functions.
For a single brane, there is no $\zeta_\a (\a \in\kern-0.8em / q_A)$,
so \p{cond1} gives
\bea
\b(u)= -\frac{\e_A a_A}{2\td}\phi(u).
\ena
Using the function $Y$ in the above coordinate transformation, $D_1$ can be gauged away.
This was also noted in Ref.~\cite{CDEG}.
On the other hand, $X(\tilde u)$ can be used to gauge away $\xi$,
or choose any function of $u$ as $\xi$.
Hence there are $(d+1)$ degrees of freedom in the present single brane system.
We note that the term $C_1$ was not considered in Ref.~\cite{CDEG}.
For a single brane, it is natural to take $\xi$ and all $\zeta_\a$ equal.
Still we have four arbitrary functions.

If we set $A_1(u)=B_1(u)=0$,
we recover our previous solutions~\cite{MOTW},
while, if we set $C_1(u)=0$, we find the solutions by Craps et al.~\cite{CDEG}.
Indeed, one can check that their solution corresponds to the specific choice
\bea
h_A(u) = e^{-f(u)},~~
\xi(u) = \zeta_\a=-\frac{11-p}{8} f(u),~~
\b(u) = \frac{p-3}{8}f(u),~~
\phi(u) = \frac{7-p}{4} f(u)
\ena
with the correspondence (left is our notation)
\newcommand{\lra}{\leftrightarrow}
\bea
q_A \lra p,~~
d-1 \lra p,~~
\td \lra 7-p,
\ena
with only one arbitrary function $f(u)$, whereas ours have four.

As an example, let us consider D3-brane ($d=4$). In this case,
$a_A=0$ and we have five arbitrary functions of $u$; $\zeta_1, \zeta_2, \phi, h_3$
and $C_1$.
If we set $\xi=\zeta_1=\zeta_2= f(u)/2$, we find
the similar solution in \cite{DMNT1,DMNT2}, although $h_3$ depends on $u$.

\subsection{Intersecting two branes}

Let us consider two intersecting branes $A$ and $B$.
In this case, \p{last_eq} gives
\bea
\left(
r^{\tilde{d}+1}K'
\right)' \hs{-2}
&=& \hs{-2}-2e^{-2(\xi-\beta)}r^{\tilde d+1}H_AH_B
\left[{\dot{h}_A\dot{h}_B\over H_AH_B}+
{\ddot{h}_A\over H_A}+{\ddot{h}_B\over H_B}+2(\dot{\beta}-\dot{\xi})
\left({\dot{h}_A\over H_A}+{\dot{h}_B\over H_B}\right)+V(u)\right].
\nn
&~&\label{m1}
\ena
Substituting \p{harm1} and sorting out the terms in the orders of $r$,
we find
\bea
\left(
r^{\tilde{d}+1}K'
\right)'
=A_2(u)r^{\tilde d+1}+B_2(u)r+{C_2(u)\over r^{\tilde d-1}}
\,,
\label{m2}
\ena
where
\bea
A_2(u)&=&-2e^{-2(\xi-\beta)}\left[
\dot{h}_A\dot{h}_B+\ddot{h}_A{h}_B+\ddot{h}_B{h}_A+2(\dot{\beta}-\dot{\xi})
(\dot{h}_A{h}_B
+\dot{h}_B{h}_A)+Vh_Ah_B
\right]
\nn
B_2(u)&=&-2e^{-2(\xi-\beta)}\left[
\ddot{h}_AQ_B+\ddot{h}_BQ_A+2(\dot{\beta}-\dot{\xi})
(\dot{h}_AQ_B
+\dot{h}_BQ_A)+V(h_AQ_B+h_BQ_A)
\right]
\nn
C_2(u)&=&-2e^{-2(\xi-\beta)}VQ_AQ_B
\label{m3}
\ena
Integrating Eq. (\ref{m2}), we find
\bea
K={A_2(u)\over 2(\tilde d+2)}r^2 -{B_2(u)\over 2(\tilde d-2)}r^{-(\tilde d-2)}
+{C_2(u)\over 2(\tilde d-2)(\tilde d-1)}r^{-2(\tilde d-1)}
-{D_2(u)\over \tilde d}r^{-\tilde d}+E_2(u)
\label{sol1}
\,,
\ena
for the case that $\tilde d$ is not equal to 1 nor 2,
where $D_2(u)$ and $E_2(u)$ are arbitrary functions.
In the cases of $\tilde d=1$ and 2, we find
\bea
K={A_2(u)\over 6}r^2+{B_2(u)\over 2}r
+C_2(u)\ln r
-{D_2(u)\over r}+E_2(u)
\label{sol2}
\,,
\ena
and
\bea
K={A_2(u)\over 8}r^2
+{1\over 2}\left(B_2(u)-{C_2\over r^2}\right)\ln r
-{C_2(u)+2 D_2(u)\over 4 r^{2}}+E_2(u)
\label{sol3}
\,,
\ena
respectively.
The latter case corresponds to an M2-M5 intersecting brane system.
If we set $A_2(u)=0$, $B_2(u)=0$ and $C_2(u)=0$,
we again recover our previous solutions~\cite{MOTW}.

In the similar way to the single brane case, we shall gauge away
$E_2$ and $\xi$ in what follows.
Compared with the single brane system, our intersecting brane system has
additional functions $h_B(u)$, but there is one additional constraint \p{cond1}
for the additional brane.
Hence we are again left with $(d+1)$ arbitrary functions in the present system.

Let us give some concrete examples. The D1-D5-brane is given by
\bea
ds^2 &=&  H^{\frac{1}{4}}_1   H^{\frac{3}{4}}_5 \Big[
  H^{-1}_1   H^{-1}_5
( -2 d  u d  v +   K(  u,  r)d  u^2)
+   H_5^{-1} \sum^4_{\alpha =1}e^{2 \zeta_\a(  u)} dy^2_\a \nn
&& \hs{20} +\; e^{2  \b( u)}(d  r^2 +   r^2 d\Omega^2_3)\Big], \nn
  \Phi &=& \ln \left(\frac{  H_1}{  H_5}\right)^{\frac{1}{2}}
+ \phi(u)
\label{d1-d5 new}
\,,
\ena
where
\bea
H_A&=&h_A(u)+{Q_A\over r^2}~~~(A=1, 5)\,,
\label{D1D5_H}
\ena
and (\ref{sol3}) without $E_2$.
It also follows from \p{cond1} that
\bea
  \b(u)=\frac14   \phi(u) =-\sum_{\a=1}^4  \zeta_\a(u).
\ena
This solution has seven arbitrary functions $h_1(u),
 h_5(u), \zeta_\a(u)$, and $D_2(u)$ in $K$, while
$A_2, B_2$, and $C_2$ in $K$ are given by Eq. (\ref{m3}) with $\xi=0$.

If we assume $V(u)=0$, \p{cond1} together with \p{def_V}
yields $\zeta_\a=\b=\phi=0$.
Then $ K(u, r)$ is found to be
\bea
  K (u, r)=\frac{\bar A_2(u)}{8}   r^2
+\frac{\bar B_2(u)}{2} \ln{ r}+ \bar D_2(u)  r^{-2}\,,
\ena
where
\bea
\bar A_2(u) &\equiv& -8 \left[h_1^{\half}
\left(\dot h_5
 h_1^{\half}\right)^{\cdot}+ h_5^{\half} \left(
\dot h_1 h_5^{\half}
\right)^{\cdot}\right]\,,\nn
\bar B_2(u)&\equiv&-8\left(
Q_5 \dot h_1+Q_1 \dot h_5\right)^{\cdot}\,,
\label{tilde K}
\ena
and $\bar D_2(u)$ is an arbitrary function of $u$.
In the case of $\bar B_2=0$, $K$ is called asymptotically Brinkmann form.
There remain three arbitrary functions of $u$; $h_1(u),
h_5(u)$ and $\bar D_2(u)$.

Next let us consider D2-D6-brane solution:
\bea
ds^2 &=& H^{\frac{3}{8}}_2  H^{\frac{7}{8}}_6 \Big[
 H^{-1}_2  H^{-1}_6
( -2 d u d v + K(u, r) du^2)
+  H^{-1}_2  H^{-1}_6 e^{2 \zeta_{1}(u)}(dy^1)^2 \nn
&& +\;  H^{-1}_6\sum^5_{\alpha =2} e^{2\zeta_{\a}(u) }dy^2_\a
+e^{2 \beta(u)}
\left(d r^2 +  r^2 d\Omega^2_2\right)\Big], \nn
\Phi &=& \frac{1}{4}\ln H_2 - \frac{3}{4}\ln  H_6 +  \phi( u)
\,.
\label{d2-d6}
\ena
In this case, there are eight nontrivial $u$-dependent functions
$h_2(u), h_6(u), D_2(u)$ and $\zeta_{\a} (\a=1, \ldots, 5)$ with
\bea
\phi(u)=\frac{4}{3}\b(u) = -\sum_{\a=2}^5 \zeta_\a(u).
\ena

\section{A fluctuating ``black hole"}

In the static case, one can construct a black hole solution from
the intersecting brane system via compactification.
Hence we may find a time-dependent black hole solution by compactifying
the present time-dependent intersecting brane systems\cite{MOU,MN,GM}.
We give a simple example of this type.

Let us consider the simple case
of D1-D5 intersecting brane system with
\bea
H_A=1+{Q_A\over r^2}~,~{\rm and}~~~K={2Q_{\rm w}(u)\over r^2}
\,,
\ena
where $Q_{\rm w}$ is a function of $u$.
This can be obtained for the choice $h_1= h_5=1$
and $\bar D_2=-2 Q_{\rm w}(u)$.
One can check $\bar A_2(u)=B_2(u)=0$ by \p{tilde K} easily.
Introducing new function $H_{\rm w}=1+K(u,r)/2$, we find the metric
\bea
ds_{10}^2
&=& H_1^{-3/4}H_5^{-1/4}\left[-H_{\rm w}^{-1} dt^2+H_{\rm w}
\left(dz+{(H_{\rm w}-1)\over H_{\rm w}}dt\right)^2  \right] \nn
&& +H_1^{1/4}H_5^{3/4}\left[ H_5^{-1}\sum_{\alpha=1}^4 dy_\alpha^2
+ dr^2+r^2 d\Omega^2_3\right],
\label{metric_D1D5}
\,
\ena
where $u=(t-z)/\sqrt{2}$ and $v=(t+z)/\sqrt{2}$.

In order to perform a compactification,
we write our metric (\ref{metric_D1D5})  as
\bea
ds_{10}^2&=&
\left(H_1^{-{1\over 12}}H_5^{5\over 12}H_{\rm w}^{-{1\over 3}}\right)
\,ds_5^2
+H_1^{-3/4}H_5^{-1/4}H_{\rm w}
\left(dz+{(H_{\rm w}-1)\over H_{\rm w}}dt\right)^2
+H_1^{1/4}H_5^{-1/4}\sum_{\alpha=1}^4 dy_\alpha^2,
\nn
&&
\label{metric_D1D52}
\,
\ena
where
\bea
ds_5^2=-\Xi_5^2dt^2
+\Xi_5^{-1}(dr^2 +r^2 d\Omega_3^2),
\label{5D_metric}
\ena
and $\Xi_5=(H_1H_5H_{\rm w})^{-1/3}$ gives
the five-dimensional metric in the Einstein frame.

All toroidal $y_\alpha$-coordinates can be compactified, but
the compactification of the $z$-coordinate is not trivial.
We have to impose a periodic condition on the metric functions, which
explicitly depend on $z$ through the $u$-coordinate.
Here we assume that the function $H_{\rm w}(u,r)$ is periodic in the $u$ direction.
As a concrete example, we choose a periodic function as
\bea
Q_{\rm w}(u)
=Q_0\left[1+\epsilon \cos \left({\sqrt{2}u\over  R}\right)\right]
\,,
\ena
where $R$ is a radius of the $z$-space and $\epsilon$
is a positive constant.
If fact, the metric is invariant under the discrete transformation of
$z\rightarrow z+2\pi nR$ ($n \in \bf{Z}$).
The explicit form of the metric function $\Xi_5$ is given by
\bea
\Xi_5=\left[\left(1+{Q_1\over r^2}\right)
\left(1+{Q_5\over r^2}\right)\left(1+{Q_0\over r^2}
\left[1+\e\cos \left({t-z\over  R}\right)\right]
\right)\right]^{-1/3}
\label{Xi}
\,.
\ena

In order to avoid a closed timelike curve, the $z$-direction must be
spacelike. This condition requires that $\epsilon \leq 1$; otherwise
$H_{\rm w}$ becomes negative at least in the limit of $r\rightarrow
0$, where we expect a horizon.
$\epsilon =1$ must be excluded because
the charge $Q_{\rm w}$ vanishes at $u=\pi R/\sqrt{2}$,
when a singularity may appear at $r=0$.
Hence we assume that $0<\epsilon<1$.

The metric (\ref{5D_metric}) with (\ref{Xi}) gives effectively
a five-dimensional time-dependent spacetime
although it also depends on the $z$-coordinate.
It describes an explicit example of
a spacetime excited by a pyrgon
which may appear in Kaluza-Klein compactification\cite{KS,GR}.
Since it is asymptotically flat,
one may define the ``mass'' of this object as
\bea
M={\pi[Q_1 +Q_5 +Q_{\rm w}(u)]\over 4G_5}
\,,
\ena
which  oscillates in time.
The surface of $r=0$ is a candidate for horizon because
it is the case when the spacetime is static ($\epsilon=0$).
Hence one may naively think that this spacetime describes
a time-dependent oscillating ``black hole''.
However the ``mass'' depends not only on time $t$ but also
on the inner space coordinate $z$.

If the compactification radius $R$ is small enough,
 we may not see $z$-dependence in a global scale.
Taking an average over the internal $z$-space, we find that
the mean mass $\langle M\rangle$ is given by
\bea
\langle M\rangle={\pi(Q_1+Q_5+Q_0)\over 4G_5}
\,.
\ena
We may also find that the ``mass" $M$ fluctuates around this average value
with the amplitude
\bea
{\sqrt{\langle(\Delta M)^2\rangle}\over \langle M\rangle}
={Q_0\epsilon \over \sqrt{2}(Q_1+Q_5+Q_0)}
\,.
\ena
and the typical frequency $\omega=\sqrt{2}/R$.

The Bekenstein-Hawking black hole entropy, which
is proportional to the horizon area, may also
fluctuate around the averaged value
\bea
\langle S\rangle={\langle {\cal A}\rangle\over 4G_5}={\pi^2
\sqrt{Q_1Q_5 Q_0}\over 2G_5}
\ena
with the amplitude
\bea
{\sqrt{\langle(\Delta S)^2\rangle}\over \langle S\rangle}
={\epsilon \over \sqrt{2}}
\,.
\ena

This spacetime describes
a five-dimensional compact object in a global scale,
but it shows fluctuations near the ``horizon" ($r=0$).
Hence it is not a deterministic five-dimensional spacetime.
We can regard it as a fluctuating ``black hole",
but $r=0$ may not be a true horizon.

Although this spacetime looks like
a fluctuating ``black hole" in five dimensions,
it is a deterministic spacetime in six dimensions.
In fact, when we approach the ``horizon",
we will see the internal compact $z$-space as well as the periodic
time dependence.
Hence the spacetime is essentially six-dimensional,
whose metric is given by
\bea
ds_6^2=\Xi_6(r)\left[-2du dv + K(u,r) du^2\right]
+ \Xi_6^{-1}(r) (dr^2 +r^2 d\Omega_3^2), \quad
\Xi_6 (r)\equiv (H_1 H_5)^{-1/2} \,.
\ena
It is obtained by compactification of all toroidal $y_\alpha$-coordinates
in ten-dimensional spacetime as
\bea
ds_{10}^2&=&H_1^{1/4}H_5^{3/4}\left[H_1^{-1}H_5^{-1}(-2du dv + K(u,r)du^2)
+ H_5^{-1}\sum_{\a=1}^4 dy_\a^2 +(dr^2+r^2 d\Omega_3^2)\right] \nn
&=&H_1^{-1/4}H_5^{1/4} ds_6^2 + H_1^{1/4}H_5^{-1/4}\sum_{\a=1}^4 dy_\a^2
\,.
\ena
Although this spacetime is compact in the $z$-direction
as well as in the toroidal $y_\alpha$-direction,
the $z$-direction is not homogeneous.
As a result, the spacetime is time-dependent but it is no longer
spherically symmetric, i.e. it depends on  $z$ as well as $t$, $r$.
The inhomogeneity in the $z$-direction
becomes prominent especially in the scale near (or smaller than)
the compactification radius $R$.
This spacetime is regular at $r=0$, which is shown by calculating
the curvature invariants. The Ricci scalar curvature
and Kretschmann invariant are given by
\bea
{\cal R} = -\frac{r^4(Q_1-Q_5)^2}{((r^2+Q_1)(r^2+Q_5))^{5/2}}
 ~~\rightarrow ~~0~~~{\rm as}~~~r\rightarrow 0
\ena
and
\bea
{\cal R}_{\mu \nu\rho\sigma}{\cal R}^{\mu\nu\rho\sigma}
&=&\frac{24Q_1^4Q_5^4+96Q_1^3Q_5^3(Q_1+Q_5)r^2
+O(r^4)}{(r^2+Q_1)^5(r^2+Q_5)^5} \nn
&\rightarrow &\frac{24}{Q_1Q_5}~~~~~~~~~~~~~~~~~~~~~~~~
{\rm as}~~~r\rightarrow 0
\ena
We thus see that these scalars do not diverge
at the ``horizon" ($r=0$).

Hence we conclude that this solution describes a static and spherically
symmetric five-dimensional compact
object with fluctuations in a large scale, but
 it becomes a periodically oscillating and non-spherical
six-dimensional object in a small scale.

There are several questions with this solution which
deserve further consideration.
Does this metric really describes a time-dependent black hole or else?
Is the horizon, if it exists, time-dependent?
How is the mass of the ``black hole" defined?
When we approach the ``horizon", what kind of spacetime structure do we see?
Those questions are interesting by themselves and are left for future study.

\section{Concluding Remarks}

We have constructed a fairly general family of supersymmetric
solutions in time- and
space-dependent wave backgrounds in supergravity theories.
These solutions describe intersecting $p$-branes
embedded into time-dependent dilaton-gravity plane waves of an arbitrary
(isotropic) profile, with the brane world-volume aligned parallel to the
propagation direction of the wave.

In our previous paper~\cite{MOTW}, we have derived this class of solutions
but restricted the wave profile by setting $K=0$ in Eq.~\p{last_eq} for
simplicity, and then solved the resulting equation.
However we have shown in this paper that if we regard \p{last_eq} as the equation
for K, we can get more general class of solutions and we have solved it explicitly.
We have also discussed how many degrees of freedom we are left with,
and found that we have $(d+1)$ arbitrary functions.
This approach has recently been taken in \cite{CDEG} for a single brane.
Our solutions here include not only single but also intersecting brane solutions
as well as wider solutions including more arbitrary functions than those in \cite{CDEG}.
To investigate intersecting branes system is very important, because
this class of solutions may describe standard model of particle physics,
higher-dimensional black holes and so on.
Thus we hope that our solutions provide a useful basis to investigate
various physical phenomena.

As a simple physical application, we have also used one of the solutions to construct
higher-dimensional time-dependent black hole. The example we have considered
is the D1-D5 intersecting brane system, and we have proposed an effectively
compactified solution in five dimensions.
The result is an ``oscillating black hole solution'' in five dimensions.
This black hole looks like five-dimensional ``oscillating'' black hole whose frequency is
$\omega=\sqrt{2}/R$ ($R$ is compact radius) if seen from infinity.
On the other hand it looks like six-dimensional black hole near ``horizon''.
This is presented just as a simple example of possible compactification of
our brane configurations.
We may consider more complicated brane configurations and also other
variants of compactification.
It would be very interesting to contemplate further applying our
solutions to more interesting (higher-dimensional) black holes.
It is also an interesting subject to study non-extreme extension of our solutions.

\section*{Acknowledgements}

We would like to thank Ben Craps, Oleg Evenin and Gary Gibbons
for useful discussions.
The work was supported in part by the Grant-in-Aid for
Scientific Research Fund of the JSPS Nos. 19540308, 20540283
and 21$\cdot$\,09225, and by the Japan-U.K. Research Cooperative Program.



\begin{thebibliography}{10}

\bibitem{time1}
H.~Liu, G.~W.~Moore and N.~Seiberg,
  JHEP {\bf 0206} (2002) 045
  [arXiv:hep-th/0204168];
  JHEP {\bf 0210} (2002) 031
  [arXiv:hep-th/0206182].
\bibitem{time2}
A.~Hashimoto and S.~Sethi,
  Phys.\ Rev.\ Lett.\  {\bf 89} (2002) 261601
  [arXiv:hep-th/0208126].
\bibitem{S}
J.~Simon,
  JHEP {\bf 0210} (2002) 036
  [arXiv:hep-th/0208165].
\bibitem{CLO}
R.~G.~Cai, J.~X.~Lu and N.~Ohta,
  Phys.\ Lett.\ B {\bf 551} (2003) 178
  [arXiv:hep-th/0210206].
\bibitem{null}
N.~Ohta,
  Phys.\ Lett.\ B {\bf 559} (2003) 270
  [arXiv:hep-th/0302140].
\bibitem{OPS}
N.~Ohta, K.~L.~Panigrahi and S.~Siwach,
 Nucl.\ Phys.\ B {\bf 674} (2003) 306
 [Erratum-ibid.\ B {\bf 748} (2006) 309] [arXiv:hep-th/0306186].
\bibitem{bb1}
B.~Craps, S.~Sethi and E.~P.~Verlinde,
  JHEP {\bf 0510} (2005) 005
  [arXiv:hep-th/0506180].
\bibitem{bb2}
M.~Li,
  Phys.\ Lett.\ B {\bf 626} (2005) 202
  [arXiv:hep-th/0506260].
\bibitem{bb3}
M.~Li and W.~Song,
  JHEP {\bf 0510} (2005) 073
  [arXiv:hep-th/0507185].
\bibitem{bb4}
Y.~Hikida, R.~R.~Nayak and K.~L.~Panigrahi,
  JHEP {\bf 0509} (2005) 023
  [arXiv:hep-th/0508003].
\bibitem{bb5}
B.~Chen,
  Phys.\ Lett.\ B {\bf 632} (2006) 393
  [arXiv:hep-th/0508191].
\bibitem{bb6}
J.~H.~She,
  JHEP {\bf 0601} (2006) 002
  [arXiv:hep-th/0509067].
\bibitem{bb7}
B.~Chen, Y.~l.~He and P.~Zhang,
  Nucl.\ Phys.\ B {\bf 741} (2006) 269
  [arXiv:hep-th/0509113].
\bibitem{IKO}
T.~Ishino, H.~Kodama and N.~Ohta,
  Phys.\ Lett.\ B {\bf 631} (2005) 68
  [arXiv:hep-th/0509173].
\bibitem{bb8}
D.~Robbins and S.~Sethi,
  JHEP {\bf 0602} (2006) 052
  [arXiv:hep-th/0509204].
\bibitem{She:2005qq}
J.~H.~She,
  Phys.\ Rev.\ D {\bf 74} (2006) 046005
  [arXiv:hep-th/0512299].
\bibitem{bb10}
B.~Craps, A.~Rajaraman and S.~Sethi,
  Phys.\ Rev.\ D {\bf 73} (2006) 106005  [arXiv:hep-th/0601062].
\bibitem{CH}
C.~S.~Chu and P.~M.~Ho,
 JHEP {\bf 0604} (2006) 013 [arXiv:hep-th/0602054].
\bibitem{bb12}
S.~R.~Das and J.~Michelson,
 Phys.\ Rev.\ D {\bf 73} (2006) 126006 [arXiv:hep-th/0602099].
\bibitem{DMNT1}
S.~R.~Das, J.~Michelson, K.~Narayan and S.~P.~Trivedi,
 Phys.\ Rev.\ D {\bf 74} (2006) 026002 [arXiv:hep-th/0602107].
\bibitem{lin}
F.~L.~Lin and W.~Y.~Wen,
 JHEP {\bf 0605} (2006) 013  [arXiv:hep-th/0602124].
\bibitem{bb13}
E.~J.~Martinec, D.~Robbins and S.~Sethi,
 JHEP {\bf 0608} (2006) 025 [arXiv:hep-th/0603104].
\bibitem{bb14}
H.~Z.~Chen and B.~Chen,
 Phys.\ Lett.\ B {\bf 638} (2006) 74 [arXiv:hep-th/0603147].
\bibitem{bb15}
T.~Ishino and N.~Ohta,
 Phys.\ Lett.\ B {\bf 638} (2006) 105 [arXiv:hep-th/0603215].
\bibitem{NP}
R.~R.~Nayak and K.~L.~Panigrahi,
 Phys.\ Lett.\ B {\bf 638} (2006) 362 [arXiv:hep-th/0604172].
\bibitem{KO}
H.~Kodama and N.~Ohta,
 Prog. Theor. Phys. {\bf 116} (2006) 295 [arXiv:hep-th/0605179].
\bibitem{BC}
B.~Craps,
  Class.\ Quant.\ Grav.\  {\bf 23} (2006) S849
  [arXiv:hep-th/0605199].
\bibitem{NPS}
R.~R.~Nayak, K.~L.~Panigrahi and S.~Siwach,
 Phys.\ Lett.\ B {\bf 640} (2006) 214
 [arXiv:hep-th/0605278].
\bibitem{OP}
N.~Ohta and K.~L.~Panigrahi,
  Phys.\ Rev.\  D {\bf 74} (2006) 126003
  [arXiv:hep-th/0610015].
\bibitem{DMNT2}
S.~R.~Das, J.~Michelson, K.~Narayan and S.~P.~Trivedi,
  Phys.\ Rev.\  D {\bf 75} (2007) 026002
  [arXiv:hep-th/0610053].
\bibitem{BPRW}
J.~Bedford, C.~Papageorgakis, D.~Rodriguez-Gomez and J.~Ward,
  Phys.\ Rev.\  D {\bf 75} (2007) 085014
  [arXiv:hep-th/0702093].
\bibitem{dual}
C.~S.~Chu and P.~M.~Ho,
  JHEP {\bf 0802} (2008) 058
  [arXiv:0710.2640 [hep-th]].
\bibitem{MN}
  K.~Madhu and K.~Narayan,
  Phys.\ Rev.\  D {\bf 79} (2009) 126009
  [arXiv:0904.4532 [hep-th]].
\bibitem{ADG}
  A.~Awad, S.~R.~Das, A.~Ghosh, J.~H.~Oh and S.~P.~Trivedi,
  Phys.\ Rev.\  D {\bf 80} (2009) 126011
  [arXiv:0906.3275 [hep-th]].
\bibitem{Narayan}
  K.~Narayan,
  arXiv:0909.4731 [hep-th].
\bibitem{MOTW}
  K.~Maeda, N.~Ohta, M.~Tanabe and R.~Wakebe,
  JHEP {\bf 0906} (2009) 036
  [arXiv:0903.3298 [hep-th]].
\bibitem{KU}
H.~Kodama and K.~Uzawa,
  JHEP {\bf 0603} (2006) 053
  [arXiv:hep-th/0512104];\\
P.~Binetruy, M.~Sasaki and K.~Uzawa,
  arXiv:0712.3615 [hep-th].
\bibitem{MOU}
  K.~Maeda, N.~Ohta and K.~Uzawa,
  JHEP {\bf 0906} (2009) 051
  [arXiv:0903.5483 [hep-th]].
\bibitem{PT}
G.~Papadopoulos and P.~K.~Townsend,
  Phys.\ Lett.\  B {\bf 380} (1996) 273
  [arXiv:hep-th/9603087].
\bibitem{T}
A.~A.~Tseytlin,
  Nucl.\ Phys.\  B {\bf 475} (1996) 149
  [arXiv:hep-th/9604035].
\bibitem{AEH}
  R.~Argurio, F.~Englert and L.~Houart,
  Phys.\ Lett.\  B {\bf 398} (1997) 61
  [arXiv:hep-th/9701042].
\bibitem{NO}
 N.~Ohta,
 Phys.\ Lett.\ B {\bf 403} (1997) 218 [arXiv:hep-th/9702164].
\bibitem{MO}
  Y.~G.~Miao and N.~Ohta,
  Phys.\ Lett.\  B {\bf 594} (2004) 218
  [arXiv:hep-th/0404082].
\bibitem{CDEG}
  B.~Craps, F.~De Roo, O.~Evnin and F.~Galli,
  JHEP {\bf 0907} (2009) 058
  [arXiv:0905.1843 [hep-th]].
\bibitem{MN} K. Maeda and M. Nozawa, arXiv 0912.2811[hep-th].
\bibitem{GM} G.W. Gibbons and K. Maeda, arXiv 0912.2809[gr-qc].
\bibitem{KS}
E.~W.~Kolb and R.~Slansky,
  Phys.\ Lett.\  B {\bf 135} (1984) 378.
\bibitem{GR}
G.~W.~Gibbons and P.~J.~Ruback,
  Phys.\ Lett.\  B {\bf 171} (1986) 390.


\end{thebibliography}
\end{document}